\documentclass[10pt,a4paper]{book}
\usepackage{frontier07}

\newcommand{\gray}{$\gamma$-ray}

\newcommand{\aap}{A\&A}
\newcommand{\apj}{ApJ}
\newcommand{\app}{Astropart.\ Phys.}
\newcommand{\icrc}{Int.\ Cosmic Ray Conf.}
\newcommand{\mnras}{MNRAS}

\newcommand{\ssr}{Space Sci.\ Rev.}
\newcommand{\pubjournal}[5]{#1 {\bf #2}, #3 (#4)}

\begin{document}
\newcounter{ctr}
\setcounter{ctr}{\thepage}
\addtocounter{ctr}{8}

\talktitle{Origin and propagation of cosmic rays (some highlights)}
\talkauthors{Igor V.\ Moskalenko \structure{}}

\begin{center}
\authorstucture[]{
Hansen Experimental Physics Laboratory and 
Kavli Institute for Particle Astrophysics and Cosmology, 
   Stanford University, Stanford, CA 94305, U.S.A.} 


\end{center}

\shorttitle{Origin and propagation of cosmic rays} 

\firstauthor{Igor V.\ Moskalenko}

\begin{abstract}

The detection of high-energy particles, cosmic rays (CRs), deep inside 
the heliosphere implies that there are, at least,
three distinctly different stages in the lifetime of a CR particle: 
acceleration, propagation in the interstellar medium (ISM),
and propagation in the heliosphere. 
Gamma rays produced by 
interactions of CRs with gas, radiation, and 
magnetic fields can be used to study their spectra in different
locations. 
Still, accurate direct measurements of CR species inside
the heliosphere (such as their spectra and abundances) are extremely
important for the understanding of their origin and propagation. 
In this paper, an emphasis is made on very recent advances 
and especially on those where GLAST and PAMELA
observations can lead to further progress in our understanding of CRs.

\end{abstract}

\section{Introduction}
Cosmic rays and \gray{s} are intrinsically connected: 
\gray\ emission is a direct probe of proton and 
lepton spectra and intensities in distant locations of the Galaxy.
Diffuse emission accounts for
$\sim$80\% of the total \gray{} luminosity of the Milky Way 
and is a tracer of interactions of CR particles in the ISM \cite{diffuse}. 
On the other hand, direct measurements of CR species can be used to 
probe CR propagation, to derive propagation parameters, and to test
various hypotheses \cite{Strong2007}. 
Luckily, two missions of the present and near future are
targeting these issues and present unique opportunities for
breakthroughs. 
The Payload for Antimatter-Matter Exploration and 
Light-nuclei Astrophysics (PAMELA) \cite{Picozza2006} has been launched 
in June 2006 and is currently in orbit. 
During its projected 3 yr lifetime it will measure light CR 
nuclei, antiprotons, and positrons 
in the energy range 50 MeV/n -- 300 GeV/n with high precision.
The Gamma-ray Large Area Space Telescope (GLAST) \cite{glast} 
is scheduled for launch in early 2008. 
It has significantly improved sensitivity, angular resolution, 
and much larger field of view than its predecessor EGRET and will 
provide excellent quality data in the energy range 20 MeV -- 300 GeV. 

\section{Cosmic-ray accelerators}
A supernova (SN) -- CR connection has been discussed since the mid-1930s 
when Baade and Zwicky proposed that SNe are responsible for the observed CR
flux.
The first direct evidence of particle acceleration 
up to very high energies (VHE) 
came from observations of synchrotron X-rays from
the supernova remnant (SNR) SN1006
\cite{Koyama1995}.
More recently, observations of TeV \gray{s}
\cite{Aharonian2006a} confirm the existence of VHE particles.
Still, definitive proof that SNRs are accelerating protons is absent. 
Recent observations of the SNR RX J1713 by HESS suggest that its spectrum
is consistent with the decay of pions produced in $pp$-interactions 
\cite{Aharonian2006a},
while the spectrum from inverse Compton scattering (ICS) does not 
seem to fit the observations. 
However, a calculation of
ICS and synchrotron emission using a one-zone model \cite{Aharonian1999}
and a new calculation of the interstellar radiation field 
(ISRF) shows that a leptonic origin is also 
consistent with the data \cite{Porter2006}. 
Another interesting case study is
the composite SNR G0.9+0.1 near the Galactic center \cite{Aharonian2005a} 
which can also be fitted using the leptonic model \cite{Porter2006}. 
In this instance, the major contribution comes from
ICS off optical photons while the \gray\ spectrum exhibits a ``universal''
cutoff in the VHE regime due to the Klein-Nishina effect. 
If this modelling is correct, GLAST observations can be used
to probe the ISRF in the Galactic center. 
Observations of SNRs
by GLAST will be vital in distinguishing between leptonic or hadronic
scenarios as their predictions for the spectral
shape in the GeV energy range are distinctly different.

A new calculation of the ISRF shows that it is more
intense than previously thought, especially in the inner Galaxy
where the optical and infrared photon density exceeds that 
of the cosmic microwave background (CMB) by a factor of 100 \cite{Porter2006}. 
For a source in the 
inner Galaxy, properly accounting for inverse Compton energy losses
flattens the electron spectrum in the source compared to the case of
pure synchrotron energy losses \cite{Hinton2006}. 
This effect
leads to a flatter intrinsic \gray\ spectrum at the source. 
On the other hand, the intense ISRF also leads to
$\gamma\gamma$-attenuation which starts at much lower energies than for
the CMB alone \cite{Moskalenko2006c}. 
For VHE \gray\ sources located in the inner Galaxy,
the attenuation effects should be seen for energies $\sim$30 TeV 
\cite{Porter2007}.

A new class of VHE \gray\ sources and thus CR accelerators, 
close binaries, has been recently found by HESS \cite{Aharonian2006b}.
The observed orbital modulation due to the
$\gamma\gamma$-attenuation on optical photons of a companion star
testifies that the VHE \gray\ emission is produced near the compact object.
Such an effect has been predicted long ago in a series of papers 
\cite{Moskalenko_binaries}, where the light curves were calculated 
for binaries which were suspected to be VHE
emitters at that time, like Cyg X-3. 
The phase of the maximum of the
emission depends on the eccentricity of the orbit and its
orientation; for orbital parameters of LS 5039 it is about 0.7 
\cite{Moskalenko_binaries}, in agreement with observations. 
For a
recent discussion of the orbital modulation in LS 5039, see
\cite{Dubus2006}. 
GLAST observations in the GeV--TeV range will be the
key to understanding the emission mechanism(s).

\section{Propagation of cosmic rays and diffuse gamma-ray emission}
\subsection{Particle propagation near the sources}

Diffuse \gray\ emission in the TeV energy range 
has been recently observed by HESS 
\cite{Aharonian2006c} from the Galactic center. 
The emission 
clearly 
correlates with
the gas column density as traced by CS. 
If this emission is associated with a
relatively young SNR, say Sgr A East, 
observation of the individual clouds will tell
us about CR propagation there. 
A simple back-of-the-envelope calculation
shows that if the SNR age is $<$10 kyr and the shock speed is
$<$$10^4$ km/s, the shell size should be $<$100 pc, while
the emission is observed from distant clouds up to 200 pc
from the Galactic center. 
The emission outside the shock,
therefore, 
has to be
produced by protons which were accelerated by the shock and left it
some time ago. 
The spectrum of such particles can be approximated by
a $\delta$-function in energy which depends on the SNR age;
the resulting \gray\ spectrum is essentially flatter than expected 
from a power-law proton spectrum in the shell (Figure~1) 
\cite{Gabici2007,Moskalenko2007b}. 
Observations of individual clouds 
in the GLAST energy range will be a direct probe of this
model and thus of proton acceleration in SNRs.

\begin{figure}
\begin{center}
\epsfig{figure=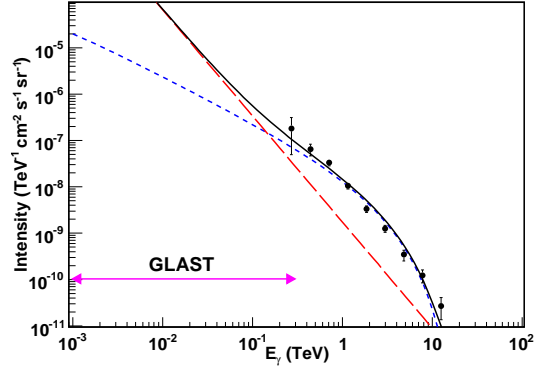,width=7cm}
\end{center}
\caption{The spectrum of \gray{s} \cite{Moskalenko2007b}
from the gas clouds outside of the SNR shell (monoenergetic protons
of 25 TeV, dots) and from the shell 
(power-law with index $-2.29$, dashes); 
normalisations are arbitrary.
The solid line is the total spectrum.
Data: HESS observations of the Galactic center ridge \cite{Aharonian2006a}.
\label{cap:gamma-ray-flux}}
\end{figure}

Milagro has recently observed the diffuse emission at 12 TeV
from the Cygnus region \cite{Abdo2007}. 
The observed emission, 
after subtraction of point sources correlates with gas column density.
The VHE \gray\ flux is found to be larger than predicted by the conventional 
and even optimized model tuned to fit the GeV excess \cite{Strong_diffuse}.
This may imply that freshly accelerated particles interact with the 
local gas, but other possibilities such as ICS
or unresolved point sources can not be excluded 
on the base of Milagro observations alone.

These observations show that the diffuse emission exists even 
at VHE energies and variations in brightness are large. 
A contribution of Galactic CRs to the diffuse
emission in this energy range is still significant. 
Observations of the
diffuse emission may be used to study CR propagation and their 
penetration into molecular clouds. 
GLAST is ideally suited to address these issues.

\subsection{Galactic cosmic rays}\label{galacticCR}
Propagation in the ISM changes the initial spectra and composition of 
CR species due
to spallation, energy losses, energy gain (e.g., diffusive reacceleration), 
and other processes (e.g., diffusion, convection). 
The destruction of primary nuclei via spallation gives rise to secondary 
nuclei and isotopes which are rare in nature (i.e., Li,
Be, B), antiprotons, pions and kaons that decay producing 
secondary leptons and \gray{s}.
Studies of stable secondary nuclei (Li, Be, B, Sc, Ti, V) allow the ratio 
(halo size)/(diffusion coefficient) to be determined and the incorporation
of radioactive secondaries 
($^{10}$Be, $^{26}$Al, $^{36}$Cl, $^{54}$Mn) is used to find the
diffusion coefficient and the halo size separately. 
For a recent review on CR propagation see \cite{Strong2007}.

Measurement of the B/C ratio with a single instrument and in a wide energy
range is long overdue. 
The best data $>$0.8 GeV/n to-date are
those taken by the HEAO-3 experiment more than 25 years ago, 
while modern spacecraft,
e.g., ACE, provide high quality data at low energies 150--450 MeV/n 
\cite{Wiedenbeck2001}. 
The sharp maximum in the B/C ratio observed at $\sim$1 GeV/n
is difficult 
to explain in a physical model and has been long debated; it may
well be an instrumental artefact. 
On the other hand, the high energy tail
of the B/C ratio is sensitive to the rigidity dependence of the
diffusion coefficient and thus its accurate measurement can be 
used to distinguish between models of CR propagation in the
ISM. 
The PAMELA has the capability of measuring the B/C ratio in the energy 
range 100 MeV/n -- 250 GeV/n and will address both issues.

The majority of CR antiprotons observed near
the Earth are secondaries produced in collisions of CRs with interstellar gas.
Because of the kinematics of this process, the spectrum of
antiprotons has a unique shape distinguishing it from other
CR species. It peaks at $\sim$2 GeV decreasing sharply
toward lower energies. 
Because of their high production threshold and the
unique spectral shape antiprotons can be used
to probe CR propagation in the ISM and the heliosphere, 
and to test the local Galactic average proton spectrum 
(for a discussion and references see 
\cite{Moskalenko_pbars}). 
Because the $pp$ (and $\bar pp$) 
total inelastic cross section is ten times smaller than
that of carbon, the ratio $\bar p/p$ can be used to derive the 
diffusion coefficient in a much larger Galactic volume than the B/C ratio.
The CR $\bar p$ spectrum may also contain signatures of exotic processes,
such as, e.g., WIMP annihilation.
However, currently available data (mostly from BESS flights \cite{bess}) 
are not accurate enough, while published estimates of the expected 
flux differ significantly; in particular, 
the reacceleration model
underproduces antiprotons by a factor of 
$\sim$2 at 2 GeV \cite{Moskalenko_pbars}.
Secondary CR positrons are produced in the
same interactions as antiprotons and are potentially able to contribute
to the same topics. 
Accurate measurements of the CR $e^+$ spectrum may also
reveal features associated with the 
sources of primary positrons, such as pulsars.
During its lifetime, PAMELA will measure
CR antiprotons and positrons in the energy range 50 MeV -- 250 GeV with 
high precision \cite{Lionetto}. 
Independent CR $\bar p$ 
measurements below $\sim$3 GeV will be provided by the new 
BESS-Polar instrument scheduled to fly in December of 2007 \cite{bess}.

The diffuse emission is a tracer of interactions of CR particles 
in the ISM and is produced via ICS, bremsstrahlung, and $\pi^0$-decay.
The puzzling excess in the EGRET data above 1 GeV \cite{Hunter1997}
relative to that expected has shown up in
all models that are tuned to be consistent with local nucleon
and electron spectra \cite{Strong_diffuse}. The excess has shown
up in all directions, not only in the Galactic plane.
If this excess is not an instrumental 
artefact, it may be telling us that the CR intensity 
fluctuates in space which could be the result of 
the stochastic nature of supernova events. 
If this is true, the local CR spectra are not
representative of the local Galactic average.
Because of the secondary origin of CR antiprotons, their intensity
fluctuates less than that of protons and $\bar p$ measurements 
can be used instead to derive the \emph{average} local intensity of CR protons.
Interestingly,
a model based on a renormalized CR proton flux (to fit antiprotons) and
a CR electron flux (using the diffuse emission itself), the so-called
optimized model, fits the all-sky EGRET data well \cite{Strong_diffuse}
providing a feasible explanation of the GeV excess. 
The GLAST
observations of the diffuse emission will be able to resolve
this puzzle. 
On the other hand, accurate measurements of CR antiprotons
by PAMELA can be used to test the CR fluctuation hypothesis.

\section{Cosmic rays in the heliosphere}
Interestingly, GLAST will be able to trace the CRs in the
heliosphere as well. 
The ICS of CR electrons off solar photons 
produces \gray{s} with a broad distribution on the sky contributing 
to a foreground that would otherwise be ascribed to the Galactic and 
extragalactic diffuse emission \cite{solarIC}.
Observations by GLAST can be used to monitor the heliosphere and determine 
the electron spectrum as a function of position from distances as large
as Saturn's orbit to close proximity of the Sun, thus enabling unique 
studies of solar modulation. 
A related process is the
production of pion-decay \gray{s} in interactions of 
CR nuclei with gas in the solar atmosphere \cite{Seckel1991}. 
The albedo \gray{s} will be observable by GLAST
providing a possibility to study CR cascade development in the solar atmosphere,
deep atmospheric layers, and magnetic field(s).
The original analysis of the EGRET data assumed that the Sun is a point source
and yielded only an upper limit \cite{Thompson1997}. 
However, a recent re-analysis of the EGRET data \cite{Orlando2007} has
found evidence of the albedo (pion-decay) and the extended ICS emission.
The maximum likelihood values appear to be consistent with the 
predictions.

GLAST will also be able to measure CR electrons directly. 
It is a very efficient electron detector able to operate
in the range between $\sim$20 GeV and 2 TeV \cite{Moiseev2007}. 
The total number of detected electrons will be $\sim$$10^7$ per year.
Accurate measurements of the CR electron spectrum are very important
for studies of CR propagation and diffuse \gray\ emission.
There is also the possibility to see the features
associated with the local sources of CR electrons.

The Moon emits \gray{s} \cite{Thompson1997,Morris1984} 
due to CR interactions in its rocky surface. 
Monte Carlo simulations of the albedo spectrum 
using the GEANT4 framework show that it is very 
steep with an effective cutoff around 3 GeV and exhibits a 
narrow pion-decay line at 67.5 MeV \cite{Moskalenko2007a}.
The albedo flux below $\sim$1 GeV significantly depends on the incident CR 
proton and helium spectra which change over the solar cycle. 
Therefore, it is possible to monitor
the CR spectrum at 1 AU using the albedo \gray\ flux. 
Simultaneous measurements of CR proton and helium
spectra by PAMELA, and observations of the albedo \gray{s} by 
GLAST, can be used to test the model predictions.
Since the Moon albedo spectrum is well understood, it can be used
as a standard candle for GLAST. 
Besides, the predicted pion-decay line at 67.5 MeV and the 
steep spectrum at higher energies present 
opportunities for in orbit energy calibration of GLAST.

\medskip
I thank Troy Porter for useful suggestions.
This work was supported in part by a NASA APRA grant.

\end{document}